\documentclass[12pt]{article}

\usepackage[english]{babel}

\usepackage{latexsym}
\usepackage{amssymb}
\usepackage{epsfig}

\begin{document}

\title{Anomalous thermal properties of a harmonic chain with correlated
isotopic disorder}

\author{ I. F. Herrera-Gonz\'{a}lez${}^{1}$, F.~M.~Izrailev${}^{2}$,
L.~Tessieri${}^{1}$ \\
{\it ${}^{1}$ Instituto de F\'{\i}sica y Matem\'{a}ticas} \\
{\it Universidad Michoacana de San Nicol\'{a}s de Hidalgo} \\
{\it Morelia, Mich., 58060, Mexico} \\
{\it ${}^{2}$ Instituto de F\'{\i}sica, Universidad Aut\'{o}noma de
Puebla,} \\
{\it Apartado Postal J-48, Puebla, Pue., 72570, Mexico}}

\date{29th March 2010}

\maketitle

\begin{abstract}
We analyse the thermal properties of a harmonic chain with weak
correlated disorder. With the use of a perturbative approach we derive
analytical expressions for the time-evolution of the chain temperature and
of the heat flow when both ends of the chain are coupled to heat baths.
Our analytical and numerical results demonstrate that specific long-range
correlations of the isotopic disorder can suppress or enhance the vibrational
modes in pre-defined frequency windows. In this way one can arrange a
frequency-selective heat flow through disordered chains.
\end{abstract}

Pacs: 44.10.+i, 63.50.Gh, 63.20.Pw

\section{Introduction}

One-dimensional (1D) harmonic chains constitute convenient models for the
effective study of thermal transport in solids. On the one hand, they
represent one of the few cases for which analytical results can be obtained
with a relative ease.
On the other hand, they can be used to describe the features of solids in the
low-temperature region, where nonlinear effects can be neglected.
For these reasons the heat conduction in 1D harmonic chains has long been a
subject of active investigation (see~\cite{Lep03,Dha08} and references
therein).

This paper is devoted to the study of the thermal properties of 1D harmonic
chains with correlated isotopic disorder.
Since the mid-1990s the interest in models of this kind has been
constantly increasing~\cite{Sou94}, especially after it became clear that
specific long-range correlations in random potentials can create a sort of
delocalisation transition even in 1D models~\cite{Mou98,Izr99}.
In the past ten years many studies, both at the
theoretical~\cite{Izr01,Izr05,Tes06} and experimental~\cite{Kuh00,Kro02}
level, have thrown light on the phenomenon of enhancement or suppression
of the Anderson localisation in low-dimensional systems with correlated
random potentials.
Within the field of 1D models with correlated disorder, some recent studies
have begun to analyse the chains with long-range correlated random masses.
These early works were focused on the anomalous diffusion of an initially
localised energy input~\cite{Mou03}, or on the structure of the eigenmodes
and on the statistics of the vibrational spectrum~\cite{Shi07}.

The present work adopts a different perspective, considering harmonic chains
as open systems weakly coupled at both ends to Langevin heat baths as
in the Casher-Lebowitz model~\cite{Leb71}.
The purpose of our study is twofold: to analyse the approach to the
steady-state regime of the chain and to discuss the effects of long-range
correlations of the isotopic disorder on the heat conduction.
Accordingly, we first analyse the dynamics of 1D harmonic chains with
random masses, without specifying the statistical properties of the disorder.
Using standard tools of stochastic calculus, we obtain analytical
expressions for the temperature profile and heat flux, describing
the evolution of the system toward the steady-state regime. In the
stationary limit, our expressions reduce to the formulae first
obtained by Matsuda and Ishii~\cite{Mat70} and Lebowitz and
coworkers~\cite{Leb71}.

After, we focus on the thermal effects of long-range correletions.
In a homogeneous chain all vibrational modes are
extended, whereas in a random chain with uncorrelated disorder only
low-frequency modes are significantly non-localised.
However, specific long-range correlations of the disorder can generate
the delocalisation or enhanced localisation of vibrational modes
in predefined frequency windows.
Thus, with an appropriate selection of the frequency intervals one can create
a random chain with a finite fraction of extended modes in the middle of
the frequency spectrum. This allows one to have a random chain with specific
properties of thermal transport peculiar to models that are neither
completely random nor perfectly ordered.

As an example, we show how to construct a random chain whose conductivity
tends to scale with the size of the system regardless of the boundary
conditions. We also demonstrate that with a proper choice of long-range
correlations one can enhance the localisation of the eigenmodes within a
wide range of the frequency spectrum. In this way one can reduce the
thermal conductivity of the chain with respect to that of a chain with
white-noise disorder.

\section{The temperature profile and heat flux}

We consider a harmonic chain of $N$ masses $m_{1},\ldots, m_{N}$ and we
study its dynamics when the chain is weakly coupled to two Langevin baths
at temperatures $T_{+}$ and $T_{-}$.
The time evolution is described by the system of stochastic It\^{o}
equations
\begin{equation}
\begin{array}{ccl}
dq_{i} & = & \displaystyle \frac{p_{i}}{m_{i}} dt \\
dp_{i} & = & \displaystyle \left[ \sum_{j} {\bf D}_{ij} q_{j} -
\delta_{i1} \frac{\lambda}{m_{1}} p_{1} -
\delta_{iN} \frac{\lambda}{m_{N}} p_{N} \right] dt \\
& + & \displaystyle
\delta_{i1} \sqrt{2 \lambda k_{B} T_{-}} \, dW_{-} +
\delta_{iN} \sqrt{2 \lambda k_{B} T_{+}} \, dW_{+} \\
\end{array}
\label{ito1}
\end{equation}
with $i=1, \ldots, N$.
In eq.~(\ref{ito1}), $p_{i}$ and $q_{i}$ represent the momentum and the
displacement from the equilibrium position of the $i$-th mass of the
chain. The symbols $W_{+}(t)$ and $W_{-}(t)$ stand for two independent
Wiener processes, while $\lambda$ represents the coupling constant of the
chain to the baths.
Finally, ${\bf D}$ is the tridiagonal force-constant matrix, with elements
${\bf D}_{ij} = 2k \delta_{i,j} - k \delta_{i,j-1} - k \delta_{i,j+1}$
for $i=2,\ldots,N-1$ and $j=1,\ldots, N$ while the first and last
rows take different forms according to the chosen boundary conditions.
For fixed boundary conditions one has
${\bf D}_{1,j} = 2k \delta_{1,j} - k \delta_{1,j-1}$
and ${\bf D}_{N,j} = 2k \delta_{N,j} - k \delta_{N,j+1}$, while for
free boundary conditions the matrix elements are
${\bf D}_{1,j} = k \delta_{1,j} - k \delta_{1,j-1}$ and
${\bf D}_{N,j} = k \delta_{N,j} - k \delta_{N,j+1}$.

To analyse the dynamics of the chain it is convenient to switch to normal
coordinates and momenta, defined by the equations
\begin{equation}
\begin{array}{ccc}
\displaystyle
Q_{i} = \sum_{k=1}^{N} e_{k}^{(i)} \sqrt{m_{k}} q_{k}, & &
\displaystyle
P_{i} = \sum_{k=1}^{N} e_{k}^{(i)} p_{k}/\sqrt{m_{k}}.
\end{array}
\label{norcor}
\end{equation}
Here $e_{k}^{(i)}$ is the $k$-th component of the $i$-th eigenvector of the
rescaled force-constant matrix
\begin{equation}
{\bf \tilde{D}} = \left( {\bf M}^{-1/2} \right)^{\rm T} {\bf D} {\bf M}^{-1/2}
\label{dmat}
\end{equation}
with eigenvalue $\omega_{i}^{2}$, i.e.,
${\bf \tilde{D}} e^{(i)} =  \omega_{i}^{2} e^{(i)}$ .
In eq.~(\ref{dmat}), the symbol ${\bf M}$ stands for the mass matrix
${\bf M}_{ij} = m_{i} \delta_{ij}$ .
As can be seen from eq.~(\ref{norcor}), the $k$-th component of the
$i$-th vector $e^{(i)}$ defines the amplitude of the oscillations of the
$k$-th mass in the $i$-th normal mode. For this reason we will refer to the
vectors $e^{(i)}$ as displacement eigenvectors.

After passing to the normal coordinates, the dynamical equations of the chain
can be written in a compact form by introducing the $2N$-component vector
${\bf X} = (Q_{1},\ldots,Q_{N},P_{1},\ldots,P_{N})$ and the $2N \times 2N$
matrix
\begin{equation}
{\bf A}_{ij} = \delta_{i, j-N} - \omega_{j}^{2} \delta_{i-N, j} -
\lambda {\bf C}_{i-N, j-N},
\label{amat}
\end{equation}
with the coupling matrix ${\bf C}$ being defined by
\begin{equation}
{\bf C}_{ij} = \left\{ \begin{array}{ccl}
\displaystyle
\frac{e_{1}^{(i)} e_{1}^{(j)}}{m_{1}} +
\frac{e_{N}^{(i)} e_{N}^{(j)}}{m_{N}} & & \mbox{ if } i,j = 1, \ldots, N \\
0 & & \mbox{ otherwise} \\
\end{array} . \right.
\label{cij}
\end{equation}
We also need to introduce the $2N \times 2$ matrix
\begin{equation}
{\bf B} = \left( \begin{array}{cc}
0 & 0 \\
\vdots & \vdots \\
0 & 0 \\
\sqrt{2 \lambda k_{B} T_{-}/m_{1}} e_{1}^{(1)} &
\sqrt{2 \lambda k_{B} T_{+}/m_{N}} e_{N}^{(1)} \\
\vdots & \vdots \\
\sqrt{2 \lambda k_{B} T_{-}/m_{1}} e_{1}^{(N)} &
\sqrt{2 \lambda k_{B} T_{+}/m_{N}} e_{N}^{(N)} \\
\end{array} \right)
\label{bmatrix}
\end{equation}
and the two-component Wiener process
${\bf W}(t) = \left( W_{-}(t), W_{+}(t) \right)$.
Then one can write the dynamical equation in the form
\begin{equation}
d{\bf X} = {\bf A} {\bf X} \; dt + {\bf B} \; d{\bf W}.
\label{ito3}
\end{equation}
Eqs.~(\ref{amat}) and~(\ref{ito3}) show that the interaction of the chain
with the bath couples the normal modes, which are not independent as would
be the case for an autonomous chain.
From eq.~(\ref{bmatrix}) one should also notice that the coupling of each
normal mode to the baths is proportional to to the first and the $N$-th
components of the corresponding displacement vector.

Given the deterministic initial condition ${\bf X}(0) = {\bf X}_{0}$, the
formal solution of the stochastic equation~(\ref{ito3}) can be expressed
in the form
\begin{equation}
{\bf X}(t) = {\bf U}(t) \left[ {\bf X}_{0} +
\int_{0}^{\infty} {\bf U}( -\tau) {\bf B} \, d{\bf W}(\tau)
\right] ,
\label{xprocess}
\end{equation}
where ${\bf U}(t) = \exp \left( {\bf A} t \right)$ is the fundamental
matrix of eq.~(\ref{ito3}).
Eq.~(\ref{xprocess}) defines a stochastic process which is Markovian,
Gaussian, and that becomes stationary over sufficiently long time-scales
provided that the eigenvalues of the matrix ${\bf A}$ have negative real
parts~\cite{Arn74}.
In other words, the stochastic process~(\ref{xprocess}) is a
Ornstein-Uhlenbeck process in the broad sense.
The Gaussian character of the process implies that its statistical
properties are completely determined once the first two moments are known.
One can easily show that the average value has the form
${\bf m}(t) = \langle {\bf X}(t) \rangle =
{\bf U}(t) {\bf X}_{0}$ ,
while the covariance matrix can be expressed as
\begin{equation}
\begin{array}{ccl}
{\bf K}_{ij}(t) & = & \displaystyle
\langle \left( {\bf X}_{i}(t) - {\bf m}_{i}(t) \right)
\left( {\bf X}_{j}(t) - {\bf m}_{j}(t) \right) \rangle \\
& = & \displaystyle
\int_{0}^{t} \left[ {\bf U}(\tau) {\bf B} {\bf B}^{\rm T}
{\bf U}^{\rm T} (\tau) \right]_{ij} d \tau .
\end{array}
\label{corrmat}
\end{equation}
Throughout this paper we use the symbol $\langle \cdots \rangle$ to denote
the average over realisations of the Wiener processes.

According to eq.~(\ref{xprocess}) the dynamics of the chain is completely
determined by the evolution operator ${\bf U}(t)$. This operator can be
evaluated perturbatively, provided that the chain is weakly coupled
to the baths, i.e.,
\begin{equation}
\lambda \ll \frac{|\omega_{i} - \omega_{j}|}{{\bf C}_{ij}}
\label{weakcoupling}
\end{equation}
for $i,j = 1, \ldots , N$. From a physical point of view, the weak-coupling
condition~(\ref{weakcoupling}) ensures the absence of resonance
effects.
Obviously, for eq.~(\ref{weakcoupling}) to be satisfied, the
eigenfrequencies $\{ \omega_{i} \}$ must be {\em non-degenerate}.
Below we assume that the condition~(\ref{weakcoupling}) is fulfilled.
For this reason, we do not apply the techniques used by Visscher to analyse
 the case of degenerated frequency spectra~\cite{Vis71}.

To obtain an expansion of ${\bf U}(t)$, we note that the latter is the
solution of the matrix equation
\begin{equation}
\dot{\bf U} = {\bf A} {\bf U}
\label{ueq}
\end{equation}
with the initial condition ${\bf U}(0) = {\bf 1}$.
Assuming that condition~(\ref{weakcoupling}) is satisfied, one can split
the matrix ${\bf A}$ in two terms
\begin{equation}
{\bf A} = {\bf A}_{0} - \lambda {\bf A}_{1} .
\label{adec}
\end{equation}
Here the first term, i.e.,
$\left[ {\bf A}_{0} \right]_{ij} =
\delta_{i, j-N} - \omega_{j}^{2} \delta_{i-N, j} - \delta_{ij} \lambda
{\bf C}_{i-N, j-N}$,
represents the generator of the unperturbed dynamics, modified by the
correction term
$\left[ {\bf A}_{1} \right]_{ij} =
\left( 1 - \delta_{ij} \right) {\bf C}_{i-N, j-N}$ .
Physically, the exact decomposition~(\ref{adec}) means that one considers
a set of $N$ damped and independent oscillators as the unperturbed
system, and regards the coupling among the oscillators as a perturbation.
The formal solution of eq.~(\ref{ueq}) can now be obtained in terms of
a time-ordered exponential,
\begin{equation}
{\bf U}(t) = \exp \left( {\bf A}_{0} t \right)
{\rm Texp} \left[ - \lambda \int_{0}^{t} {\bf A}_{1}^{(I)}(\tau) d \tau
\right] ,
\label{evop}
\end{equation}
with
${\bf A}_{1}^{(I)}(t) = \exp \left( -{\bf A}_{0} t \right) {\bf A}_{1}
\exp \left( {\bf A}_{0} t \right)$
being the perturbative term in the interaction representation.
Under the weak-coupling hypothesis~(\ref{weakcoupling}), one can
expand the time-ordered exponential in eq.~(\ref{evop}) and obtain
\begin{equation}
{\bf U}(t) = e^{{\bf A}_{0} t} - \lambda
\int_{0}^{t} e^{{\bf A}_{0} \left( t - \tau \right)}
{\bf A}_{1} e^{{\bf A}_{0} \tau} d \tau +
O \left( \lambda^{2} \right).
\label{uexp1}
\end{equation}

The knowledge of the correlation matrix~(\ref{corrmat}) enables one to
compute the temperature profile of the chain, defined via the relation
\begin{equation}
k_{B} T_{i} = \Big\langle \frac{p_{i}^{2}}{m_{i}} \Big\rangle =
\sum_{k=1}^{N} \sum_{l=1}^{N} e_{i}^{(k)} e_{i}^{(l)}
\langle P_{k}(t) P_{l}(t) \rangle .
\label{tp}
\end{equation}
For the sake of simplicity, in what follows we restrict our attention
to the case in which the masses of the chain are at rest in their equilibrium
position at the initial time, i.e., we assume that
${\bf X}_{0} = {\bf 0}$.
Inserting the momentum-momentum correlator corresponding to this initial
condition in eq.~(\ref{tp}), one eventually obtains that the temperature of
the $i$-th mass of the chain is described by the following formula,
\begin{equation}
\begin{array}{cl}
T_{i}(t) & \displaystyle
= \sum_{k} \left[ e_{i}^{(k)} \right]^{2}
\frac{\frac{T_{-} \left[ e_{1}^{(k)} \right]^{2}}{m_{1}} +
\frac{T_{+} \left[ e_{N}^{(k)} \right]^{2}}{m_{N}}}
{\frac{\left[ e_{1}^{(k)} \right]^{2}}{m_{1}} +
\frac{\left[ e_{N}^{(k)} \right]^{2}}{m_{N}}}
\left( 1 - e^{-2 \gamma_{k} t } \right) \\
& \displaystyle
+ \sum_{k,l}
\frac{\lambda e_{i}^{(k)} e_{i}^{(l)}}{\Omega_{k} + \Omega_{l}}
\left[ \frac{T_{-} e_{1}^{(k)} e_{1}^{(l)}}{m_{1}}
+ \frac{T_{+} e_{N}^{(k)} e_{N}^{(l)}}{m_{N}} \right] \\
& \displaystyle
\times e^{- \left( \gamma_{k} + \gamma_{l} \right) t}
\sin \left[ \left( \Omega_{k} + \Omega_{l} \right) t \right] \\
& \displaystyle
+ \sum_{k \neq l}
\frac{\lambda e_{i}^{(k)} e_{i}^{(l)}}{\Omega_{k} - \Omega_{l}}
\left[ \frac{T_{-} e_{1}^{(k)} e_{1}^{(l)}}{m_{1}}
+ \frac{T_{+} e_{N}^{(k)} e_{N}^{(l)}}{m_{N}} \right] \\
& \displaystyle
\times e^{- \left( \gamma_{k} + \gamma_{l} \right) t}
\sin \left[ \left( \Omega_{k} - \Omega_{l} \right) t \right] +
O \left( \lambda^{2} \right) . \\
\end{array}
\label{tempro}
\end{equation}
Here we have introduced the notations
$\gamma_{k} = \lambda {\bf C}_{kk}/2$ for the damping coefficients and
$\Omega_{k} = \sqrt{\omega_{k}^{2} - \gamma_{k}^{2}}$ for the damping-shifted
eigenfrequencies of the chain. Note that in the limit $t \to \infty$
eq.~(\ref{tempro}) reproduces the Matsuda-Ishii formula for the stationary
temperature profile.

We now turn our attention to the heat flux. Applying the standard rules
of stochastic calculus, one obtains that the average value of the energy
$H^{(k)} = 1/2 \left( P_{k}^{2} + \omega_{k}^{2} Q_{k}^{2} \right)$
associated to the $k$-th vibrational mode evolves in time according to the
equation
\begin{equation}
\frac{d \langle H^{(k)} \rangle}{dt} = J_{+}^{(k)}(t) - J_{-}^{(k)}(t)
\end{equation}
where the incoming modal heat flow is given by
\begin{equation}
\begin{array}{l}
\displaystyle
J_{+}^{(k)}(t) =
\lambda k_{B} \left( T_{+} - T_{-} \right)
\frac{\left[ e_{1}^{(k)} \right]^{2} \left[ e_{N}^{(k)} \right]^{2}}
{m_{N} \left[ e_{1}^{(k)} \right]^{2} + m_{1} \left[ e_{N}^{(k)} \right]^{2}} \\
\displaystyle
+ \lambda k_{B} \frac{\left[ e_{N}^{(k)} \right]^{2}}{m_{N}}
\frac{T_{-} \frac{\left[ e_{1}^{(k)} \right]^{2}}{m_{1}} +
T_{+} \frac{\left[ e_{N}^{(k)} \right]^{2}}{m_{N}}}
{\frac{\left[ e_{1}^{(k)} \right]^{2}}{m_{1}} +
\frac{\left[ e_{N}^{(k)} \right]^{2}}{m_{N}}}
e^{-2 \gamma_{k} t} + O \left( \lambda^{2} \right) , \\
\end{array}
\label{kinflow}
\end{equation}
while the outgoing modal heat flow is
\begin{equation}
\begin{array}{l}
\displaystyle
J_{-}^{(k)}(t) =
\lambda k_{B} \left( T_{+} - T_{-} \right)
\frac{ \left[ e_{1}^{(k)} \right]^{2} \left[ e_{N}^{(k)} \right]^{2}}
{m_{N} \left[ e_{1}^{(k)} \right]^{2} + m_{1} \left[ e_{N}^{(k)} \right]^{2}} \\
\displaystyle
-\lambda k_{B} \frac{\left[ e_{1}^{(k)} \right]^{2}}{m_{1}}
\frac{T_{-} \frac{\left[ e_{1}^{(k)} \right]^{2}}{m_{1}} +
T_{+} \frac{\left[ e_{N}^{(k)} \right]^{2}}{m_{N}}}
{\frac{\left[ e_{1}^{(k)} \right]^{2}}{m_{1}} +
\frac{\left[ e_{N}^{(k)} \right]^{2}}{m_{N}}}
e^{-2 \gamma_{k} t} + O \left( \lambda^{2} \right) . \\
\end{array}
\label{koutflow}
\end{equation}
Note again that eqs.~(\ref{kinflow}) and~(\ref{koutflow}) reduce to the
Matsuda-Ishii formulae in the asymptotic limit $t \to \infty$.

\section{Correlated disorder and structure of normal modes}

The above results were obtained for a chain composed of $N$ arbitrary
masses $m_{1}, \ldots, m_{N}$.
We now focus our attention on the case of weakly random masses, defined by
the condition that the fluctuations $\delta m_{i}$ of the masses
around their common mean value $M = \overline{m_{i}}$ be small,
i.e., $\sqrt{\overline{ \left( \delta m_{i} \right)^{2}}} \ll M$.
Here and below the symbol $\overline{( \cdots)}$ denotes the average over
realisations of the disorder.
The weak-disorder assumption allows us to use a perturbative approach 
and to work within the second-order approximation in the parameter
$\sqrt{\overline{ \left( \delta m_{i} \right)^{2}}}/M $. In this scheme, it
is enough to specify the statistical properties of the random succession
$\{ \delta m_{i} \}$ by giving the first two moments
$\overline{\delta m_{i}} = 0$ and
$\overline{\left(\delta m_{i} \right)^{2}} = \sigma^{2}$
and by assuming that the normalised binary correlator
\begin{equation}
\chi(l) = \frac{\overline{\delta m_{i} \delta m_{i+l}}}
{\overline{(\delta m_{i})^{2}}}
\label{bincor}
\end{equation}
is a known function. We assume that disorder is spatially invariant
in the mean; therefore, the binary correlator~(\ref{bincor})
depends only on the index difference $l$. We also suppose that $\chi(l)$
is a decreasing and even function of $l$.

When the chain is not coupled to the external baths, its dynamics obeys
the equation $m_{n} \ddot{q}_{n} = \sum_{k} {\bf D}_{nk} q_{k}$.
Taking the Fourier transform of both sides with respect to time, this
equation can be cast in the form
\begin{equation}
q_{n+1} + q_{n-1} + \frac{\omega^{2}}{k} \delta m_{n} \; q_{n} =
\left( 2 - \omega^{2} \frac{M}{k} \right) q_{n} .
\label{dyneq1}
\end{equation}
For weak disorder, the frequencies of the vibrational modes follow the
same dispersion law of a homogeneous chain, i.e.,
$\omega \left( \kappa \right) =
\omega_{\rm max} \left| \sin \left( \kappa/2 \right) \right|$ with
$\kappa$ being the mode wavenumber and $\omega_{\rm max} = \sqrt{4k/M}$
the largest frequency.

Eq.~(\ref{dyneq1}) has the form of the stationary Schr\"{o}dinger
equation for the 1D Anderson model.
One can use this analogy to jump to the conclusion that, in an infinite
chain, the inverse localisation length of the vibrational modes is
given by the expression
\begin{equation}
l^{-1}(\kappa) = \frac{\sigma^{2}}{2 M^{2}}
\tan^{2} \left(\frac{\kappa}{2}\right) \; W(\kappa)
\label{lyapmu}
\end{equation}
where
\begin{equation}
W(\kappa) = 1 + 2 \sum_{l=1}^{\infty} \chi(l) \cos \left( 2 l \kappa \right)
\label{ps}
\end{equation}
is the power spectrum of the random succession
$\{\delta m_{n} \}$~\cite{Izr99}.
Making use of the dispersion law, the inverse localisation
length~(\ref{lyapmu}) and the power spectrum~(\ref{ps}) can be expressed
as functions of the frequency $\omega$ rather than of the wavenumber $\kappa$.
As can be seen from eq.~(\ref{lyapmu}), the inverse localisation length
vanishes whenever the power spectrum $W(\kappa)$ is zero. Specific long-range
correlations of the disorder correspond to the power spectra which vanish in
continuous frequency intervals: this observation is the key to obtain
delocalisation transitions (within the second-order approximation) in
1D disordered models~\cite{Izr99}.

\begin{figure}[htp]
\begin{center}
\caption{Inverse entropic localisation length $l_{N}^{-1}$ versus
$\omega_{k}/\omega_{\rm max}$. The inset shows the entropic localisation length
$l_{N}$ versus $\omega_{k}/\omega_{\rm max}$.
The data were obtained for a chain
with $N=500$ masses and disorder strength $\sigma^{2}/M^{2} = 0.01$.}
\label{loclen2}
\epsfig{file=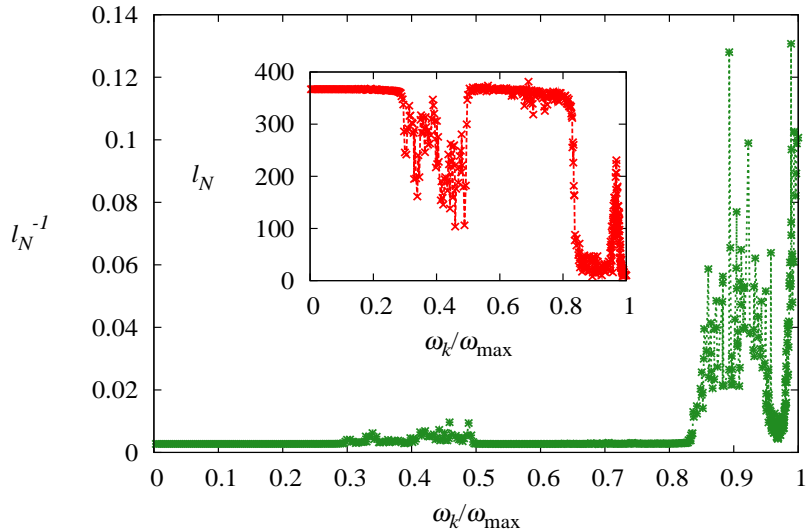,width=4.5in,height=3in}
\end{center}
\end{figure}

The theoretical formula~(\ref{lyapmu}) for the inverse localisation length
is valid only for an infinite chain, but the delocalisation transition which
it predicts persists also in finite chains with $N \gtrsim 300$ masses, as
shown by numerical computations.
This can be appreciated in Fig.~\ref{loclen2}, which represents the inverse
entropic localisation length of the vibrational modes as a function of
frequency. One can clearly see two windows of localised modes separated by
the intervals of extended modes.
The entropic localisation length, first applied in~\cite{Izr88} to
quantum chaos, is a measure of the effective number of the eigenvector
components significantly different from zero.
Specifically, the inverse entropic localisation length for an eigenmode
with displacement vector $e^{(i)}$ is defined as
$l_{N}(\omega_{i}) = \exp \left[ S_{N}(\omega_{i}) \right]$
where $S_{N}(\omega_{i}) = - \sum_{k=1}^{N} \left[ e_{k}^{(i)}\right]^{2}
\log  \left[ e_{k}^{(i)}\right]^{2}$
is the Shannon entropy associated to the $i$-th mode.
For localised eigenmodes, the entropic localisation length is proportional
to the localisation length computed in the infinite-chain limit.
Fig.~\ref{loclen2} shows the inverse entropic localisation length for
a random chain for which the binary correlator~(\ref{bincor}) has the
form
\begin{equation}
\chi(l) = \frac{1}{2 \left( \kappa_{2} - \kappa_{1} \right) l} \left[
\sin \left( 2 \kappa_{2} l \right) - \sin \left( 2 \kappa_{1} l \right) \right]
\label{lrc}
\end{equation}
exhibiting the power-law decay typical of long-range correlated disorder.

\section{Anomalous transport properties of the chain with correlated
disorder}

\begin{figure}[htp]
\begin{center}
\caption{Average modal fluxes $\overline{J_{k}}/J_{\rm max}$ versus
normalised frequency $\tilde{\omega}_{k}$ (fixed boundary conditions).
$J_{\rm max}$ is the largest modal flux.}
\label{modflux_fixbc}
\epsfig{file=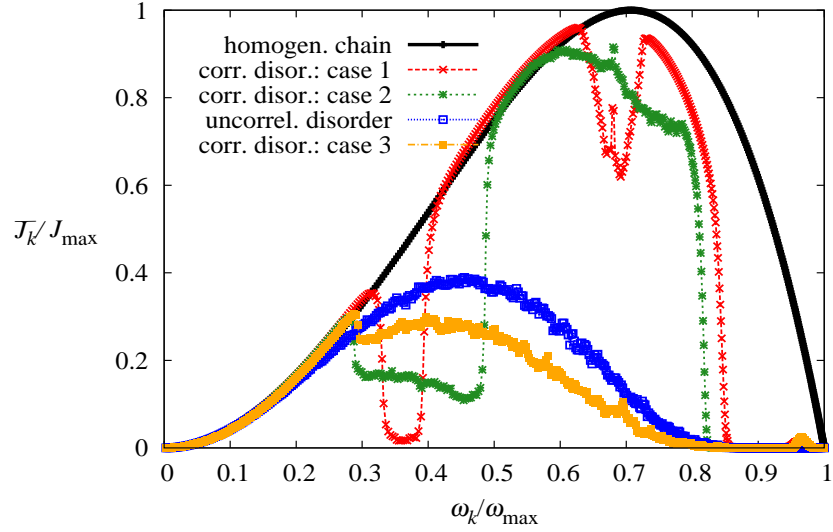,width=4.5in,height=3in}
\end{center}
\end{figure}

\begin{figure}[htp]
\begin{center}
\caption{Average modal fluxes $\overline{J_{k}}/J_{\rm  max}$ versus
normalised frequency $\tilde{\omega}_{k}$ (free boundary conditions).
$J_{\rm max}$ is the largest modal flux.}
\label{modflux_freebc}
\epsfig{file=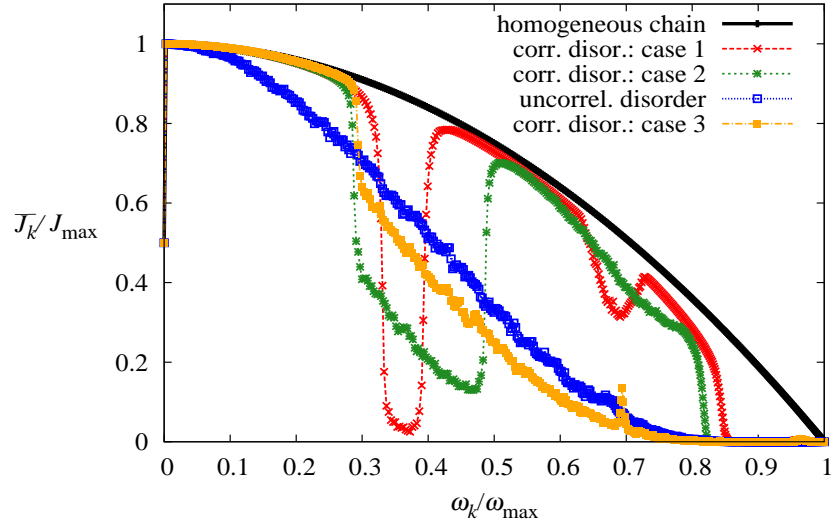,width=4.5in,height=3in}
\end{center}
\end{figure}

We can now analyse how the delocalised modes affect the heat
transport along the chain. The effect is best understood in terms of the
modal heat flows~(\ref{kinflow}) and~(\ref{koutflow}).
In figs.~\ref{modflux_fixbc} and~\ref{modflux_freebc} we show the average
modal flows in the stationary regime for fixed and free boundary conditions
(the average was taken over 1000 disorder realisations).
In both figures we show the modal fluxes for several kinds of harmonic
chains: a homogeneous chain, a chain with uncorrelated disorder, and
three chains with long-range correlated disorder. The latter cases of
correlated disorder, which we refer to as case 1, 2, and 3, have localised
modes in two windows $[\omega_{1}: \omega_{2}]$ and
$[\omega_{3}: \omega_{4}]$, and extended modes in the complementary frequency
intervals.
In case 1, the limits of the localisation windows are
$\tilde{\omega}_{1} = 0.35, \, \tilde{\omega}_{2} = 0.4, \,
\tilde{\omega}_{3}  \simeq 0.916, \, \tilde{\omega}_{4} \simeq 0.937$,
where $\tilde{\omega}_{i}= \omega_{i}/\omega_{\rm max}$. 
In case 2 we enlarge the windows of localised modes, setting their limits at
$\tilde{\omega}_{1} = 0.3, \, \tilde{\omega}_{2} = 0.5, \,
\tilde{\omega}_{3} \simeq 0.866, \, \tilde{\omega}_{4} \simeq 0.954$.
Finally, in case 3 we merge the localised-mode intervals by setting
$\omega_{2} = \omega_{3}$; we thus generate a unique window
$[\omega_{1},\omega_{4}]$ of localised modes with edges
$\tilde{\omega}_{1} = 0.3, \, \tilde{\omega}_{4} \simeq 0.954$.

In all cases the chain consists of $N=500$ masses and the disorder strength
is $\sigma^{2}/M^{2} = 0.01$.
As can be seen from figs.~\ref{modflux_fixbc} and~\ref{modflux_freebc},
the heat transport is effectively suppressed in the windows of localised
modes, being close to that of the homogeneous chain in the windows
of extended modes.
By an appropriate selection of the windows of localised and extended modes,
one can either enhance (as in case 1 and 2) or suppress (as in case 3) the
heat flow as compared with the case of uncorrelated disorder.

The delocalisation of normal modes induced by long-range correlations of
the isotopic disorder has repercussions on the conductivity of the random
chain. The most striking consequence is that, by increasing the number of
delocalised modes in the medium-to-high frequency regions, one can obtain
random chains which behave more and more as homogenous chains.
In particular, the conductivity of such random chains scales as
$\varkappa \propto N$, regardless of the chosen boundary conditions.

To understand this effect, let us focus on the case in which long-range
correlated disorder produces extended modes in a frequency window
$[\omega_{2}, \omega_{3}]$.
In the limit $N \gg 1$, the total heat flow can be approximated with the
sum of the modal fluxes~(\ref{kinflow}) over the frequency window
$[\omega_{2},\omega_{3}]$.
Assuming that the delocalised eigenmodes are not substantially altered
with respect to the homogeneous case (a conjecture confirmed by
numerical simulations for the weak-disorder case), one obtains that the
heat flow in the stationary regime is
\begin{equation}
J_{\rm cd} \simeq \frac{\lambda k_{B} (T_{+} - T_{-})}{2M}
\sum_{k = k_{2}}^{k_{3}} \left[ e_{1}^{(k)} \right]^{2} \sim
\frac{\lambda k_{B} (T_{+} - T_{-})}{2M} .
\label{totflow}
\end{equation}
By dividing the flow~(\ref{totflow}) for the temperature gradient
$(T_{+} - T_{-})/N$, one obtains that the effective conductivity of such a
chain scales as $\varkappa_{\rm cd} \sim N$, as is the case for a homogeneous
chain, and in contrast with the behaviour of a totally random chain.
For the latter, heat transport is essentially due to the $O(N^{1/2})$
low-frequency modes and one has $\varkappa_{\rm ud} \sim N^{-1/2}$
for fixed boundary conditions and $\varkappa_{\rm ud} \sim N^{1/2}$ for free
boundary conditions~\cite{Lep03}.

Our predictions are confirmed by the numerical data.
To analyse the scaling law of the conductivity, we assumed an
asymptotic behaviour of the form $\varkappa = A N^{\alpha}$ with $\alpha$
and $A$ constants.
We numerically computed the conductivity $\varkappa$ for chains of length
ranging from $N = 200$ to $N = 1000$ (with disorder strength
$\sigma^{2}/M^{2} = 0.01$).
By fitting the data for $\log \varkappa$ versus $\log N$, we obtained
values close to one for the exponent $\alpha$. Specifically, in the case
of fixed boundary conditions we respectively obtained
$\alpha_{\rm fix}^{(1)} = 0.93 \pm 0.01$ and
$\alpha_{\rm fix}^{(2)} = 0.84 \pm 0.02$ for the cases 1 and 2 of
correlated disorder.
In the same cases, for free boundary conditions, the fitting values were
$\alpha_{\rm free}^{(1)} = 0.96 \pm 0.01$ and
$\alpha_{\rm free}^{(2)} = 0.87 \pm 0.01$.
These results show that, when long-range correlations delocalise a finite
fraction of medium-to-high frequency modes (as in case 1 and 2), the
enhanced conductivity does scale as in the case of a homogeneous chain.
Especially in case 1, with a larger number of delocalised modes, the
critical exponent takes values very close to unity and is essentially
independent from the boundary conditions. 

\begin{figure}[htp]
\begin{center}
\caption{Time evolution of the temperature $T_{c}$ of the central
atom of a random chain versus $\omega_{\rm max} t$ (log-scale).}
\label{appreq}
\epsfig{file=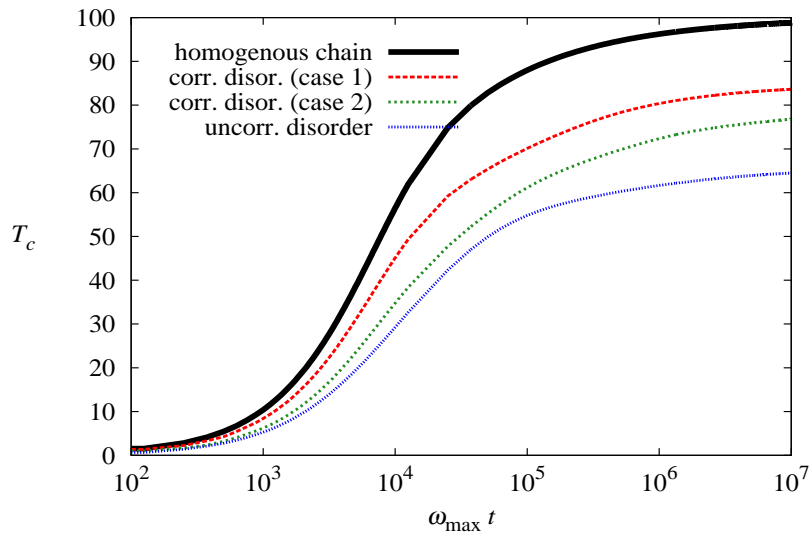,width=4.5in,height=3in}
\end{center}
\end{figure}

We would like to stress that the enhancement or reduction of the conductivity
manifests itself in the time domain, through the change of the speed with
which the system relaxes to equilibrium or approaches the steady-state
regime. By increasing the number of delocalised modes in the medium-to-high
frequency region, one can significantly speed up the approach to the
steady-state condition.
This can be seen in fig.~\ref{appreq} which represents the evolution of the
temperature of the central mass of a chain of $N=501$ atoms connected to
two baths at the temperature $T = 100$.
All the masses of the chains were initially at rest in their equilibrium
positions, and the evolution of the temperature was computed using
eq.~(\ref{tempro}).
The data shown in Fig.~\ref{appreq} were obtained for the coupling
constant $\lambda = 0.01$ and disorder strength $\sigma^{2}/M^{2} = 0.01$.
Fixed boundary conditions were chosen; the results, however, turn out to
be qualitatively the same for free boundary conditions.

\section{Conclusions}

This work analyses the heat conduction in a random chain coupled to thermal
baths. We focus on the problem of how specific long-range correlations 
of the disorder influence the thermal properties of the model.
We show that long-range correlations of the disorder can delocalise a
finite fraction of vibrational modes in the middle region of the frequency
spectrum even for chains of relatively modest length ($N \gtrsim 300$).
These delocalised modes modify both the approach to the stationary
regime and the steady-state thermal properties of the chain.
The larger the number of extended modes, the faster the chain tends
to reach the steady-state condition and the closer the thermal properties
are to those of a homogeneous chain.
In particular, when the conductivity is enhanced it exhibits anomalous
scaling $\varkappa \propto N^{\alpha}$ with $\alpha \sim 1$, independently
of the boundary conditions.
The long-range correlations of the disorder can also be used to strengthen
the localisation of the chain eigenmodes: in this way one obtains a
chain which is a better thermal insulator than a random chain with
uncorrelated isotopic disorder.

I.F.H.-G. and L.T. gratefully acknowledge the support of the CONACyT grant
No. 84604 and of the CIC-2009 grant (Universidad Michoacana).
The work of F.M.I. was partly supported by CONACyT grant No. 80715.

\end{document}